\documentclass[usenatbib,useAMS,a4paper]{mn2e}
\usepackage{epsfig}
\usepackage{txfonts,graphicx,amssymb,rotating}

\newlength{\colwidth}
\def\msun{{\rm M_{\odot}}}

\title [Self-similar disc winds]
{ A self-similar  solution for thermal disc winds}
\author[Clarke \& Alexander]{C.J. Clarke$^{1,}$\thanks{E-mail:cclarke@ast.cam.ac.uk} and 
R.D. Alexander$^{2}$ \\ 
$^1$ Institute of Astronomy, Madingley Rd, Cambridge, CB3 0HA, UK \\  
$^2$ Department of Physics \& Astronomy, University of Leicester, Leicester, LE1 7RH, UK}

\date{Replacing version published 2016: MNRAS 460,3044 correcting typographical errors}

\pagerange{\pageref{firstpage}--\pageref{lastpage}} \pubyear{2016}

\begin{document}
\def\lta{\mathrel{\spose{\lower 3pt\hbox{$\mathchar"218$}}
     \raise 2.0pt\hbox{$\mathchar"13C$}}}
\def\gta{\mathrel{\spose{\lower 3pt\hbox{$\mathchar"218$}}
     \raise 2.0pt\hbox{$\mathchar"13E$}}}
\def\Msun{{\rm M}_\odot}
\def\msun{{\rm M}_\odot}
\def\Rsun{{\rm R}_\odot}
\def\Lsun{{\rm L}_\odot}
\def\19{GRS~1915+105}
\label{firstpage}
\maketitle

\begin{abstract}
We derive a  self-similar  description for the 2D streamline topology and
flow structure of an axi-symmetric, thermally driven wind originating
from a disc in which the density is a power law function of radius. Our
scale-free  solution is strictly only valid in the absence of gravity or
centrifugal support; comparison with 2D hydrodynamic simulations of
winds from Keplerian discs  however demonstrates that the scale-free solution
is  a good approximation also in the outer regions of such discs, and 
can provide
a reasonable description even for launch radii well within  the
gravitational radius of the flow. Although other authors have considered
the flow properties along streamlines whose geometry has been specified
in advance, this is the first  isothermal calculation in which the flow geometry
and variation of flow variables along streamlines is determined self-consistently.
It is found that the flow trajectory is 
  very sensitive to the power-law index of radial density variation
in the disc: the steeper the density gradient, the  stronger is the
curvature of streamlines close to the flow base that is required 
in order to maintain momentum balance perpendicular to 
the flow. Steeper disc density profiles are also associated with more
rapid acceleration, and a faster fall-off of  density, with height above
the disc plane. The derivation of a set of simple governing equations 
for the flow structure of thermal winds from the outer
regions of power law discs offers
the possibility of deriving flow observables without having to 
resort to hydrodynamical simulation.
\end{abstract}


\section{Introduction}
   Thermally driven disc winds play an important role in the evolution of
a variety of astrophysical systems from AGN (Begelman, McKee \& Shields 1983)
to X-ray binaries (Luketic et al 2010) to protoplanetary discs (e.g.
Johnstone, Hollenbach \& Bally 1998, Alexander et al 2006, Owen et al 2010).
In particular, such winds -- where heating is provided by ultraviolet or
X-ray radiation from the young star -- are widely believed to provide an important
mechanism for clearing out proto-planetary discs and thus drawing to a close
the epoch of planet formation  (see Alexander et al 2014 for a
recent review). There is obviously considerable  interest
in  seeking observational diagnostics of such winds (Font et al 2004,
Alexander 2008, Gorti \& Hollenbach 2008. Hollenbach \& Gorti
2009, Ercolano \& Owen 2010, Owen et al 2010, Owen et al 2013). These
studies are based on numerical radiation-hydrodynamics simulations since -- even
in the simplest case of an isothermal wind with a prescribed density
structure across its base -- no {\it analytic} models for the streamline
topology and two dimensional flow structure have been available.

  Various authors have attempted to study the structure of thermally
driven disc winds. The common approach has been to {\it assume} a given
streamline structure (e.g. Begelman et al 1983, Fukue 1989, Takahara et al
1989, Fukue \& Okada 1990, Waters \& Proga 2013). In this case, not only is the variation of
cross-sectional area along a streamline bundle well defined but so also
are the external forces provided by  the gravity of the central star
and the centrifugal acceleration associated with the flow of
angular momentum
conserving disc material. In this case, if a barotropic equation
of state is assumed, the problem is a variant (with external forces)
of the `de Laval nozzle' flow of compressible fluid along pipes of
variable cross-section: there is a unique choice of flow velocity at
the base which ensures that the flow makes a transition between
subsonic and supersonic flow at
its critical point (this latter being defined by a critical
relationship between the local streamline divergence and the external
forces; Parker 1958). Although such an approach permits a consistent
solution {\it along} each streamline it does {\it not} ensure a situation
of hydrodynamical equilibrium perpendicular to the streamlines. In
general such calculations do not consider this issue since they impose
a two-dimensional streamline structure. An exception is Fukue \& Okada (1990)
who constructed a streamline topology  for which the components
of the external forces (i.e. the gravitational and centrifugal terms)
 normal to the streamline
always cancel. {\footnote{ Icke (1981) adopted a similar approach to deriving the topology of
radiatively driven winds.}}.  In fact we will show here that these terms play a minor
role in the equilibrium perpendicular to the streamlines even at radii
that are well within the `gravitational radius' ($=GM_*/c_s^2$ for stellar mass
$M_*$, sound speed $c_s$)  where the
depth of the gravitational potential well at the flow base
exceeds its  thermal energy.  This can be broadly understood in that near
the flow base the centrifugal and  gravitational terms are nearly balanced,
whereas at large radii both terms (though unbalanced) become
individually small in magnitude. We find that instead the effect that
determines the streamline structure over a wide range of launching radii is
the balance between the relevant components of the convective derivative of the velocity ($u . \nabla u$) and the pressure gradient. To put it
another way, the local {\it curvature} of the streamlines is
jointly determined by the flow velocity and the acceleration
provided by pressure gradients normal to the streamlines. At a heuristic
level this accounts for the changes in streamline topology as the density profile along the flow base is varied, an effect that is obviously
missed by formulations that instead impose the streamline structure
{\it a priori}.

  In this paper we present new similarity solutions for isothermal
flow from a disc where the density along the flow base is a power law
of radius. This similarity solution is valid in the limit of large
launching radius where we can neglect  external forces (gravity
and centrifugal terms)  and therefore differs from previously dicussed
(magneto-)hydrodynamical self-similar wind solutions which instead impose a constant
ratio of sound speed (and Alfven speed) to Keplerian speed at the flow base 
(e.g. Blandford \& Payne 1982, Contopolous \& Lovelace 1994, Li 1995, Ostriker 1997,
Ferreira \& Casse 2004). We are motivated to instead study  the globally isothermal
case, since this is a reasonable  approximation to the results of radiation
hydrodynamical modeling of disc photoevaporation from both ionising ultraviolet radiation
(Richling \& Yorke 1997) and X-rays (Owen et al 2012). Although we might expect
that neglect of external forces would result in our similarity solution  being
 valid only at large  radii,  
 we will show by comparison with  two-dimensional
isothermal hydrodynamical simulations that the flow approximately follows the
similarity solution down to launching radii as small as  $0.5 R_g$ 
(for particular power law choices).
%, and that we can understand the
%limitations of its validity by examining the magnitude of the
%terms omitted with those retained in the similarity solution. 
Section 2 sets out the derivation of the similarity solution and
Section 3 discusses its properties.  Section 4 describes the 2D hydrodynamical
solutions while Section 5 compares the  self-similar solution with
the hydrodynamic results both with and without centrifugal/gravitational
terms.  Section 6 summarises
 the properties of the solutions and their
utility for those modeling the observational consequences of disc winds.

\begin{figure}
\begin{center}
\includegraphics[width=\columnwidth,angle=270]{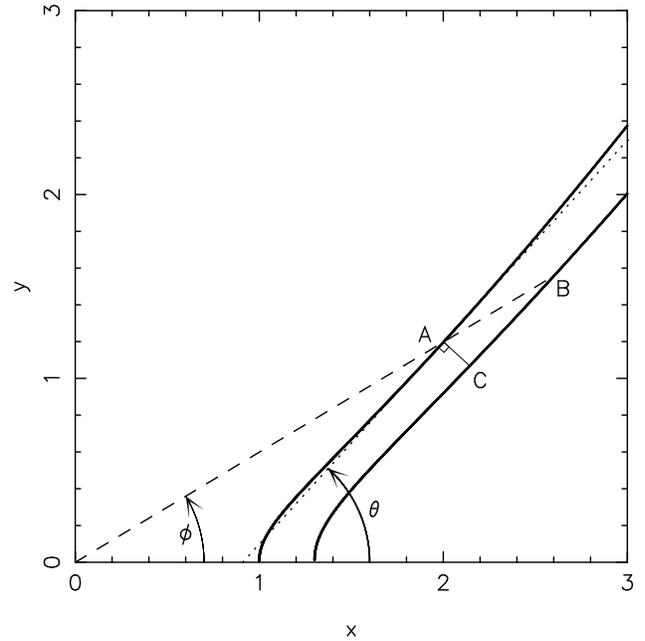}
\caption{Schematic of a pair of scale free streamlines. The component of the
pressure gradient normal to the streamline (in direction $\bf{\hat l}$)
is derived from the pressure difference between point C and point A.
This can be calculated by considering the pressure difference between A and B
(where B is the point on the adjacent streamline with the same value of $\tilde s$
as point A) and then considering the pressure difference along the adjacent
streamline between points B and C. We define $\phi$ as the polar angle with respect to the $x$-axis, and $\theta$ as the angle between the local streamline tangent (the dotted line) and the $x$-axis.}
\end{center}
\end{figure}

\section{Similarity solutions for zero-gravity, isothermal disc winds}
\subsection{Preliminaries}

We consider an axi-symmetric disc wind in the limit of large radius ($R >>
R_g =  G M/2 c_s^2$)
and where we thus omit gravitational and centrifugal force terms.
In the case that the density at the streamline base is a power law
$\rho_b \propto R_b^{-b}$ we see that there are no characteristic length
scales associated with the problem. We therefore expect the flow to
be self-similar. This means that all streamlines are simply scaled
versions of each other and thus (at the same value of $\tilde s$, the
ratio of the distance along the streamline $s$ to base radius $R_b$) all spatial
variables are a given
multiple of $R_b$; likewise the density is a given multiple of the density
at the streamline base, $\rho_b$, 
and the velocity is a given multiple of the flow velocity at the flow
base ($u_b$). We thus write:

\begin{eqnarray}
u(\tilde s,R_b) &=& u_b \tilde u(\tilde s) \\
\rho(\tilde s,R_b) &=& \rho_b \tilde \rho(\tilde s) \\
%r(\tilde s,R_b) &=& \tilde r(\tilde s) R_b \\
%y(\tilde s, R_b)&=& \tilde y(\tilde s) R_b \\
%x(\tilde s,R_b)&=& \tilde x(\tilde s) R_b \\
R_{eff}(\tilde s, R_b) &=& \tilde R_{eff} (\tilde s) R_b 
\end{eqnarray} 
 where 
%$r,x,y$ are conventional spherical polar and Cartesian
%coordinates and  
$R_{eff}$ is the local radius of curvature
of the streamline.  In principle the solution that we derive
(with gravitational and centrifugal forces omitted) would apply to
a purely 2D flow and in what follows we adopt Cartesian coordinates
$x$ and $y$ with the flow launched at $x=R_b,y=0$. Self-similarity then
implies 
\begin{eqnarray}
%u(\tilde s,R_b) &=& u_b \tilde u(\tilde s) \\
%\rho(\tilde s,R_b) &=& \rho_b \tilde \rho(\tilde s) \\
%r(\tilde s,R_b) &=& \tilde r(\tilde s) R_b \\
y(\tilde s, R_b)&=& \tilde y(\tilde s) R_b \\
x(\tilde s,R_b)&=& \tilde x(\tilde s) R_b \\
r(\tilde s,R_b) &=& \tilde r(\tilde s) R_b 
%R_{eff}(\tilde s, R_b) &=& \tilde R_{eff} (\tilde s) R_b \\
\end{eqnarray} 

\noindent  where $r^2 = x^2 + y^2$. When we compare this solution to the case
of the disc wind (with gravity and rotation included) $x$ and $y$ can
be equated with $R$ and $z$ of a cylindrical coordinate system. 
We additionally define two angles: $\phi(\tilde s)$
is the polar angle with respect to the x-axis and $\theta (\tilde s)$ 
is the angle between the local streamline tangent and the x-axis.
Figure 1 depicts two adjacent streamlines separated by $\Delta R_b$
at the base and thus (given the self-similar geometry), $\Delta r/r = \Delta R_b/R_b$. The area of a streamline bundle normalised to its value at the base is thus given by:

\begin{equation}
\tilde A = \tilde r ^2 {\rm sin} (\theta - \phi) {\rm cos} \phi
\end {equation}
(assuming that the flow is launched perpendicularly from the $z=0$ plane: see 
Section 2.2). 
 We can then  write the condition  of constant mass flux along the
streamline as

\begin{equation}
\tilde r^2 \tilde \rho \tilde u {\rm{sin}} (\theta - \phi) {\rm cos} \phi = 1
\end{equation}
Likewise the invariance of the Bernoulli stream function for an
isothermal flow in the absence of gravity or rotation can be written:

\begin{equation}
\tilde \rho \exp \biggl( {{u_b^2}\over{2 c_s^2}}(\tilde u^2 -1)\biggr) = 1 
\end{equation}
We see immediately from equation (9) that in order for the self-similarity
assumption to be valid (i.e. in order that all the scaled  quantities
are independent of streamline), $u_b$ is independent of streamline.

 Normal to the streamlines, force balance between the effect of the
pressure gradient perpendicular to the streamline (i.e. in the
${\bf \hat l}$ direction) and the relevant component of the
convective derivative of the velocity gives:

\begin{equation}
{{\tilde u^2 u_b^2}\over{\tilde R_{eff}}} = c_s^2 \nabla {\rm {ln}} \rho . \hat l
\end{equation}
where we define $R_{eff} > 0$ as implying  a geometry that is
locally convex upwards 
%CJC correctionMay 2016
(see Appendix).

In order to evaluate the right hand side of equation (10) we need to
decompose the
change in ln $\rho$ along $\hat l$ into two contributions:
the change from streamline to streamline at constant $\tilde s$
, i.e., B to A in Figure 1 (which simply relates to the change in
$\rho_b$ between the streamlines) and the change from C to B
which relates to the gradient in density along a streamline
(and which also depends on local streamline geometry).  Thus
equation (10) becomes

\begin{equation}
{{\tilde u^2 u_b^2}\over{\tilde R_{eff}}} = c_s^2 \biggl({{b}\over {{\rm{sin}} (\theta - \phi)\tilde r}} 
 + {{\partial {\rm{ln}} \rho}\over{\partial \tilde s}} {\rm cot}(\theta-\phi)\biggr)
\end{equation}
Note that the first term dominates near the base of the flow and is
positive for an outwardly decreasing density gradient ($b > 0$). The
coefficient of the second term is negative and this term dominates in
magnitude at large radii where streamlines are nearly radial.

 In order to close equations (8),(9) and (11) we need a further relationship
between the density gradient along the streamline and the local radius
of curvature. We develop this relationship using Cartesian
coordinates  with independent coordinate $\tilde y = y/R_b$ such that the streamline and
its local gradient are described in terms of $\tilde x(\tilde y)$,
  $\tilde x ^\prime (\tilde y)$. In Cartesians we can
write:

\begin{equation}
{{1}\over{\tilde R_{eff}}} = {{\tilde x ^{\prime \prime}}\over {(1 + \tilde x ^{\prime 2})^{1.5}}} 
\end{equation}
and can express $\tilde A$ (equation (7)) as: 

\begin{equation}
\tilde A = {{\tilde x^2 - \tilde x \tilde y \tilde x ^\prime}\over {(1 + \tilde x ^{\prime^2})^{0.5}}}
\end {equation} 
Then differentiation of (13) wrt $\tilde y$ yields:

\begin{equation}
\tilde x ^{\prime \prime }= {{(1 + \tilde x ^{\prime ^2})(\tilde x-\tilde y \tilde x ^\prime)\tilde x ^\prime}\over{\tilde x (\tilde y + \tilde x \tilde x ^\prime)}} - {{(1+\tilde x ^{\prime 2})^{3/2} \tilde A ^\prime} \over{\tilde x (\tilde y + \tilde x \tilde x ^\prime)}}
\end{equation}
  Combining (7)-(9) and differentiating with respect to
$\tilde y$ also yields:

\begin{equation}
\tilde A ^\prime =    \biggl({{u_b^2}\over{c_s^2}}-{{1}\over{ \tilde u^2}}\biggr) \tilde u ^\prime {\rm exp} \left( {{u_b^2}\over{2 c_s^2}}\left(\tilde u^2 -1\right)\right) 
\end{equation}
Equations (12),(14) and (15) then together allow $\tilde R_{eff}$ to 
be related to $\tilde u^\prime$  (for given $\tilde x, \tilde y, \tilde x^\prime$ and
$\tilde u$). Then using (9) to express the density gradient on the
right hand side of equaiton (11) in terms of $\tilde u^\prime$ we can convert
equation (11) into an equation for $\tilde u^\prime$ 
in terms of $\tilde x, \tilde x^\prime, \tilde x^{\prime \prime}$ and $\tilde u$:
\begin{equation}
f(\tilde x,\tilde y, \tilde x^\prime, \tilde u)\tilde u^\prime  = g(\tilde x, \tilde y, \tilde x^\prime, \tilde u)
\end{equation}
where
\begin{equation}
f =  {{- u_b^4 \tilde u^2}\over{c_s^2 \tilde x  (\tilde x \tilde x ^\prime + \tilde y)}}  {\rm exp} \left( {{u_b^2}\over{2 c_s^2}}\left(\tilde u^2 -1\right)\right) \left(1-{{c_s^2}\over{u_b^2 \tilde u^2}}\right) + {{\tilde u u_b^2 (\tilde x \tilde x ^\prime + \tilde y)}\over{(1+\tilde x ^{\prime 2})^{1/2} (\tilde x - \tilde y \tilde x ^\prime)}} 
\end{equation}

and
\begin{equation}
g= {{b c_s^2(1+\tilde x ^{\prime 2})^{1/2}}\over{(\tilde x - \tilde y \tilde x ^\prime)}}- {{u_b^2 \tilde u^2 \tilde x ^\prime (\tilde x - \tilde y \tilde x ^\prime)}\over {(1+\tilde x^{\prime 2})^{1/2}\tilde x (\tilde x \tilde x ^\prime + \tilde y )}}
\end{equation}
  Note that equation (15) is the usual expression for a de Laval nozzle, in which
the velocity structure can be computed for known variation of cross-section
along the streamline and which shows that a sonic transition is associated with
a singular point where the cross-section attains a local extremum. Naturally
the streamline solutions that we compute have this property. We however solve  
 (16) instead of (15) and find that for certain ranges of
$u_b$,   (16) admits solutions that extend to arbitrarily large radii 
without passing through a critical point. 
 This means that,
unlike the case where the variation of cross-section is specified in advance, 
there is not a {\it unique}  value of the flow velocity at the streamline base 
which allows the solution to undergo a sonic transition (although there
is a range of $u_b$ values for which the flow solution does not extend to
infinity with $\tilde u \prime$ remaining finite). Within the allowed
range of $u_b$,  we will find
solutions each of which has a different variation of
$\tilde A$ along the streamline (and a different topology),
the geometrical properties of the flow self-adjusting so as
to maintain momentum balance perpendicular to the streamlines. 
%Finally  we assume (in order to compare directly with numerical
%simulations that make this assumption e.g. Font et al 2004) that
%the flow leaves the disc perpendicularly ($\theta = \pi/2).

\subsection{Method of solution}

 We start by adopting a trial value of $u_b$ and construct the
streamline from its base ($\tilde s = 0$, $\tilde r = 1$, $\phi = 0$,
$\theta = \pi/2$). We assume that the flow leaves the disc perpendicularly (in order to compare directly with numerical
simulations that make this assumption; e.g., Font et al 2004, and the simulations
presented in Sections 4 \& 5)).  We solve for the streamline structure as an initial value problem,
 choosing  $\tilde y$ as the independent variable that is advanced along the streamline. 
At any point, P, on the streamline, at which we know the current values
of $\tilde x, \tilde y, \tilde u$ and $\tilde x ^\prime$, we use equation (16)
to evaluate $\tilde u ^\prime$;  advancing $\tilde y$
 by $\Delta \tilde y$ we then  calculate the value of $\tilde u$
at the next position along the streamline, P', 
using a first order Euler method 
 (verifying that the resulting solutions  are independent of $\Delta \tilde y$). 
Equations (14) and (15) are then used to calculate $\tilde x^{\prime \prime}$ 
at P.
The $\tilde x$ coordinate of P' and local streamline gradient
$\tilde x ^\prime$ are then readily determined:

\begin{equation}
\tilde x|_{P'} = \tilde x|_P + \tilde x^\prime |_{P}+ \Delta \tilde y + 0.5 \tilde x ^{\prime \prime}|_P  \Delta \tilde y^2
\end{equation}

\begin{equation}
\tilde x^\prime|_{P'} = \tilde x^\prime|_P + \tilde x^{\prime \prime}|_P \Delta \tilde y
\end{equation}
The streamline geometry and flow velocity are now known at point P', and the solution is then integrated to the next streamline  point. 

\section{Results}
\subsection{General properties of the flow}
We consider solutions for which $\tilde x =1, \tilde y =0$ and $\tilde x ^\prime = 0$
at  the flow base. In this case the limiting value of $\tilde u^\prime$ near the flow base is  
$b  \tilde y/({\cal M_{\rm b}}^2 (1 - {\cal M_{\rm b}}^2))$, where ${\cal M_{\rm b}}$ is the
Mach number at the flow base, so that for subsonic launch velocities the flow  accelerates 
for finite $\tilde y$. 
At large radius, the flow becomes increasingly radial
[i.e. $(1  - \tilde y \tilde x ^\prime/\tilde x)$ tends to $0$] so that the
second term in $g$  can be neglected and the limiting form of $g$ is
\begin{equation}
%g = b {{c_s^2}\over{(\tilde x - \tilde y \tilde x^{\prime})}} (1+\tilde x^{\prime 2})^{1/2}
g= {{b c_s^2(1+\tilde x ^{\prime 2})^{1/2}}\over{(\tilde x - \tilde y \tilde x ^
\prime)}}
\end{equation}

% The limiting form of $f$ is:
%\begin{equation}
%f={u_b^2 {\bigl(1-{{u_b^2 \tilde u^2 }\over{c_s^2}}\bigr) {\rm exp} \bigl({{u_b^2}\over{2c_s^2}}(\tilde u^2-1)\bigr)}\over{\tilde x (\tilde x \tilde x^\prime + \tilde y ^\prime))}} +{{\tilde u u_b^2(\tilde x \tilde x ^\prime + \tilde y)}\over{(\tilde x - \tilde y \tilde x^{\prime})(1+\tilde x^{\prime 2})^{1/2}}}
%\end{equation}

 The first term in $f$ (equation (17)) is negative in the supersonic regime
whereas $g$  and the second term in $f$ are both  
positive. Thus, depending on the value of $u_b$ and the resulting
streamline topology, the two terms in $f$  may or may not
cancel at finite $\tilde x$. If they do {\it not}, then $\tilde u^\prime$
(equation 16) remains finite and positive at all $\tilde x$ (i.e. the flow accelerates
monotonically to arbitrarily large velocity). However, if the first
term in $f$  ever becomes greater or equal in
magnitude to the second term, then $\tilde u^\prime$ becomes infinite
and changes sign. We are here concerned with the former class of
solution as representing a physical flow to infinity and we thus require that
$f$  always remains positive. We cannot impose this as
an analytic condition without solving for the streamline topology. We
nevertheless see that because the (negative) magnitude of the first
term of f  is an increasing function of $u_b$, we expect that
physical solutions that reach infinity 
 are those with relatively low $u_b$. We will
find below that this is indeed the case: for each value of $b$ we are
able to attain a range of flow solutions corresponding to a range of
$u_b$ values up to a maximum value $u_b=u_{b_{max}}(b)$. We will go on
to show in Section 5  that time-dependent hydrodynamical sumulations
in fact tend to the flow solutions with $u_b=u_{b_{max}}(b)$.   
 
\subsection{Flow solutions as a function of b}
\begin{table}
\centering
\caption{  Self-similar streamline properties. Columns  (1): index of power law for base density. (2): Maximum value of Mach number at launch such that
solution accelerates monotonically to large radius. The following properties
correspond to the streamline solution at this maximum launch Mach number:
(3) and (4) are normalised coordinates of the sonic point,  (5) is the
normalised flow
velocity at a height of $5 \times$ the initial launch radius above the
disc plane and (6) is the angle between streamline and x-axis at this location.} 
\begin{tabular}{rlllll}
\hline
\hline
${\rm b}$ & u$_{\rm b}/{\rm c}_{\rm s}$ & $\tilde {\rm x}_{\rm sonic}$ & ${\rm \tilde y}_{\rm sonic}$ & $\tilde {\rm u}/{\rm c}_{\rm s}|_{ \tilde y = 5}$ & $\theta|_ {\tilde y = 5}$\\
$0.5$ & $0.92$ & $1.02$ & $0.30$ & $ 1.92$ & $81 \deg$ \\
 $0.75$ &  $0.85$ &  $1.06$ &  $0.33$ &  $ 2.02$ & $72 \deg$ \\
$1$ & $0.77$ & $1.09$ & $0.35$ & $2.35$ & $ 76 \deg$ \\ 
$1.5$ & $0.56$ & $1.17 $ & $0.30 $ & $2.71$ & $57 \deg$ \\
 $2$ & $0.29$ & $1.23$  &  $ 0.16$ & $ 3.28$ & $38 \deg$ \\ 
\hline
\end{tabular}
\end{table}

 We detail the properties of the streamline solution as a function
of $b$ in Table 1, in each case using the solution for which the
Mach number has  the maximum value for which $f$ 
(equation (17)) remains positive (and hence $\tilde u$ increases
monotonically along the streamline). We plot the corresponding
 self-similar streamline geometries (as derived in Section 2)
as the red curves in Figure 2. It is immediately obvious that
the  flow geometry is a sensitive function of $b$, with much more
vertical trajectories being associated with lower values of $b$. This
result can be readily understood inasmuch as the value of $b$ controls the
acceleration experienced perpendicular to the streamline; for larger values
of $b$, momentum balance is achieved by the streamline  adopting a smaller
radius of curvature (equation (12)). The maximum value of flow launch
velocity also varies systematically with $b$, but more mildly, so that
the mass flux for given local base density is reduced by about a factor two
going from $b=0.5$ to $b=1.5$. 
   
\begin{figure}
\begin{center}
\includegraphics[width=\columnwidth, angle=270]{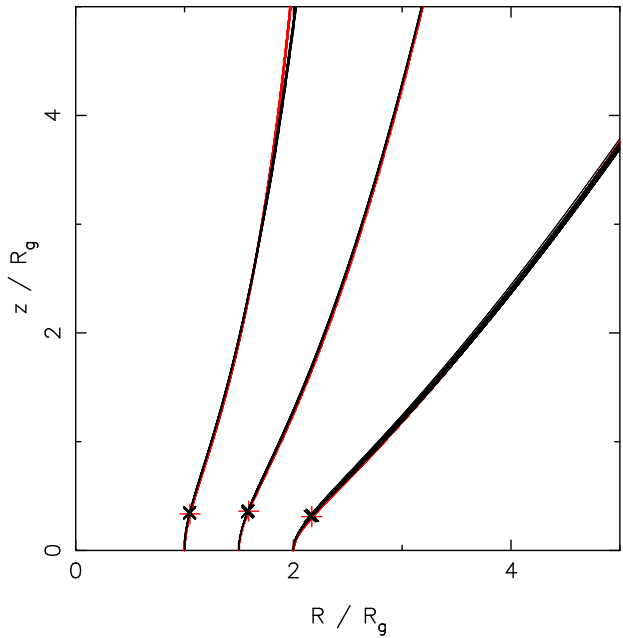}
\caption{Streamline topology for  $b=0.75$, $b=1.$ and $b=1.5$ (left to right):  self-similar  
solution (red) and scale free hydrodynamical simulation (black)  for
streamlines originating at $R=1$ (for
 clarity, the latter two streamlines are each laterally displaced
by $0.5$ while preserving the relative scale on the two axes). A series of different (re-scaled) streamlines are plotted for each hydrodynamical simulation, showing that the simulations are indeed scale-free (though slight departures from self-similarity are visible for $b=1.5$). For each streamline the sonic point is plotted as either a red ``plus'' ( self-similar solution) or a black cross ( hydrodynamical} simulations).
\end{center}
\end{figure}

%%%%%%%% NEW TEXT BY RDA, FEB 2016 %%%%%%%%%
\section{2D hydrodynamical simulations: method}

\begin{figure*}
\begin{center}
\includegraphics[width=\columnwidth,angle =270]{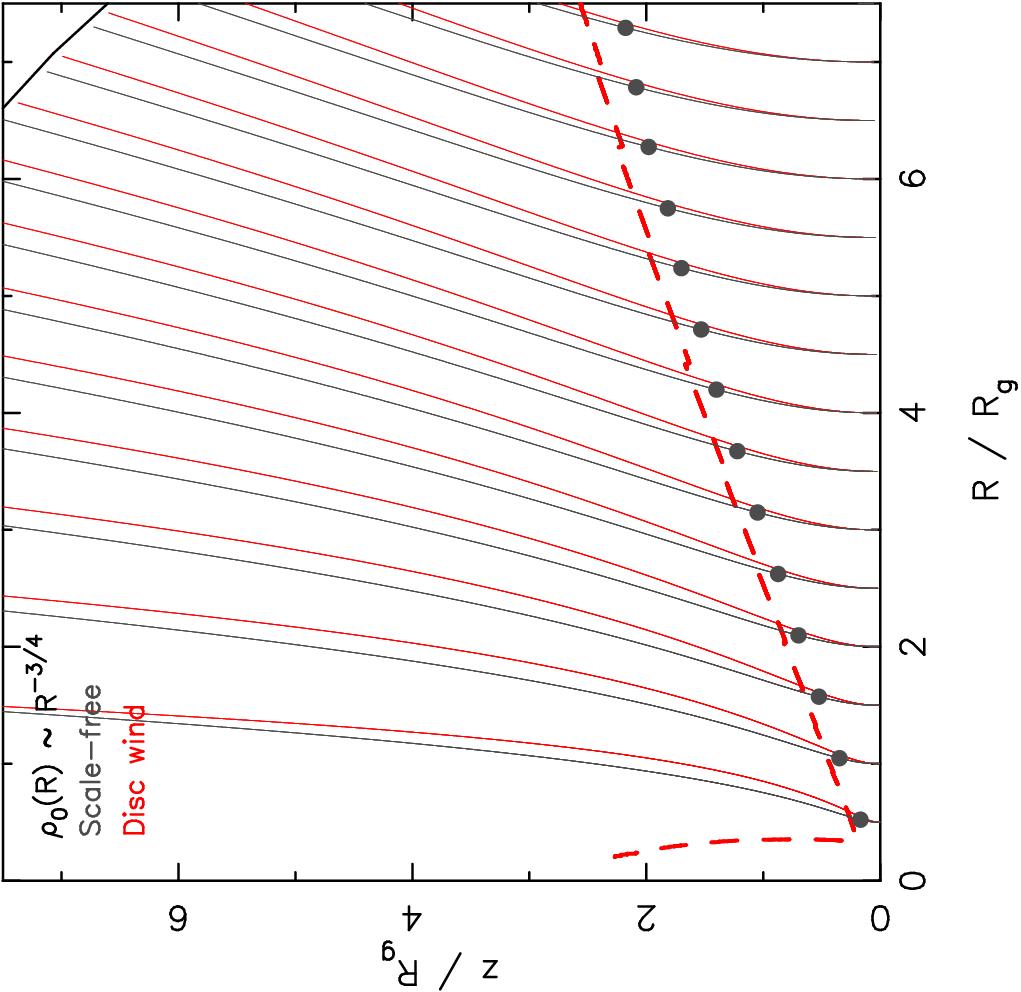}
\includegraphics[width=\columnwidth,angle=270]{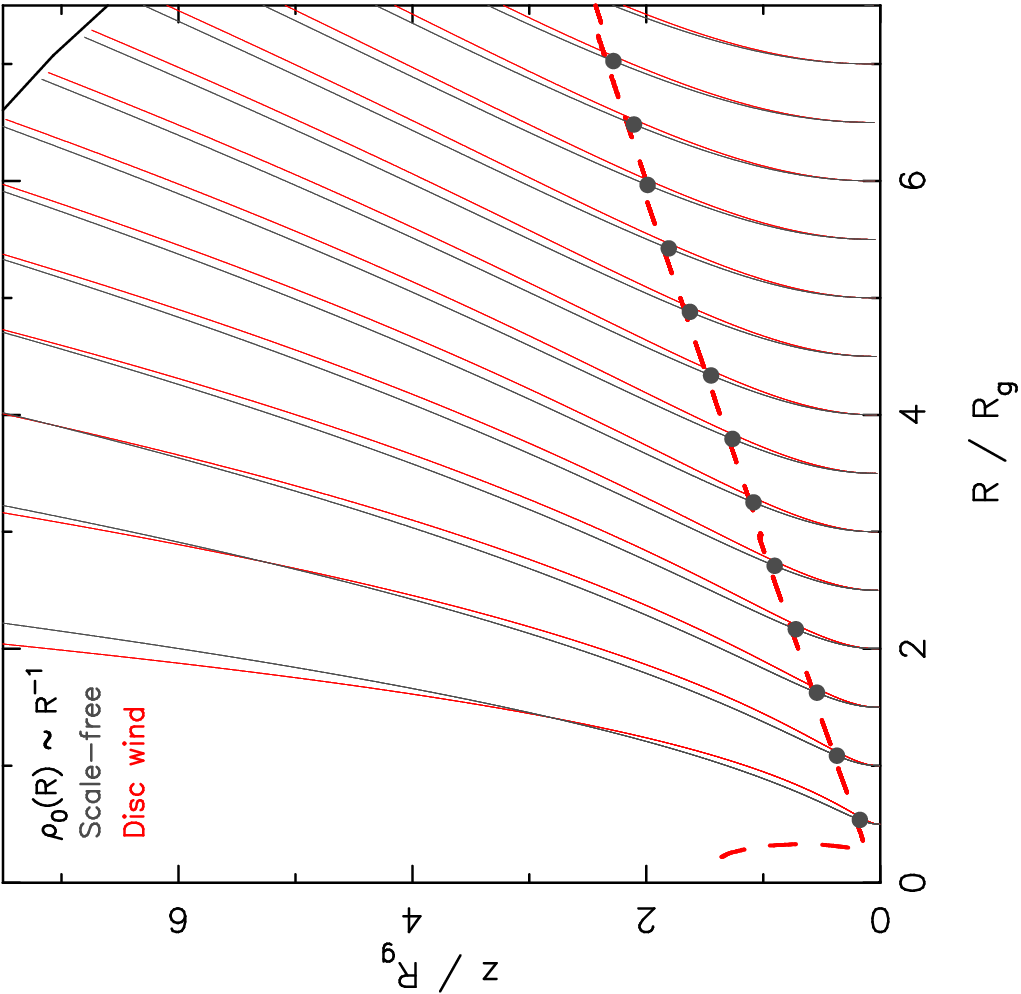}
\vspace*{12pt}

\includegraphics[width=\columnwidth,angle=270]{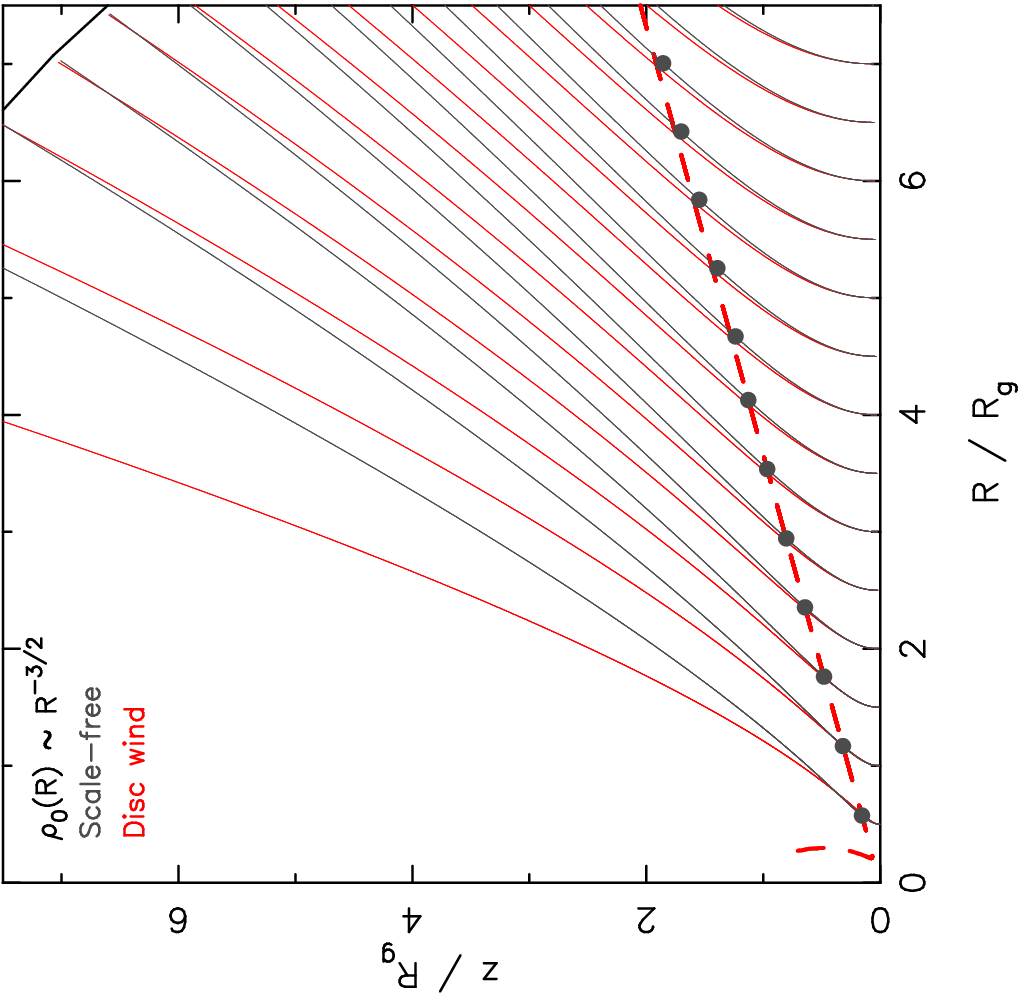}
\caption{ Comparison between the scale free (grey) and disc wind solutions
(red) for $b=0.75$, $b=1$ and $b=1.5$. The solid lines show streamlines originating from $R=0.5 R_g$, 1.0$R_g$\ldots$7.0 R_g$. The sonic surfaces in the disc wind simulations are plotted as dashed red lines, while the corresponding sonic points along the scale-free streamlines are denoted by grey circles. The solid black lines denote the boundaries of the computational grid.}
\end{center}
\end{figure*}

In order to test our  self-similar  solution we have run a series of numerical hydrodynamical calculations for comparison. We use the {\sc zeus2d} hydrodynamics code (Stone \& Norman 1992), parallelised (for a shared-memory architecture) using the OpenMP formalism\footnote{See {\tt http://openmp.org}}, and adopt the same numerical approach previously used by Font et al (2004) and Alexander (2008). We assume azimuthal and midplane symmetry, as in the  self-similar  solution, and use a polar [$(r,\theta)$] grid spanning $\theta = [0,\pi/2]$. The (fixed) grid is logarithmically spaced in $r$ and linearly spaced in $\theta$, so that the grid cells are approximately square throughout (i.e., $\Delta r = r \Delta \theta$). The grid has $N_{\theta} = 200$ cells in the polar direction and spans the range $r=[0.01R_{\mathrm g},10.0R_{\mathrm g}]$, and therefore $N_r = 883$ cells in the radial direction. We adopt the standard second-order (van Leer) interpolation scheme, and the von Neumann \& Richtmyer artificial viscosity (with $q_{\mathrm {visc}} = 2.0$). The gas has an isothermal equation of state ($P=c_{\mathrm s}^2 \rho$), and we adopt outflow boundary conditions at both the inner and outer radial boundaries. At the upper polar boundary (the $z$-axis) we adopt a reflective boundary condition, but little or no material reaches this boundary so this has no influence on the flow solutions. At the lower polar boundary ($z=0$) we impose a power-law density profile 
\begin{equation}
\rho_0(R) = \rho_{\mathrm g} \left(\frac{R}{R_{\mathrm g}}\right)^{-b} 
\end{equation}
and set the radial velocity $v_{\mathrm r}(R) = 0$ in the boundary cells. The polar velocity out of the base cells is not prescribed, but rather computed self-consistently by the hydrodynamic code. We work in dimensionless units: the unit of length is $R_{\mathrm g}$; the unit of time is the orbital period at $R_{\mathrm g}$; and the density is normalised such that $\rho_{\mathrm g} = 1$. Each model rapidly evolves towards a steady state. We run each simulation for $t=50$ time units and, to minimise numerical noise, take the average density and velocity fields over $t=[40,50]$ as the final flow solution. All simulations were run on the ALICE\footnote{See {\tt http://go.le.ac.uk/alice}} and DiRAC2/{\it Complexity}\footnote{See {\tt http://www.dirac.ac.uk}} high-performance computing clusters at the University of Leicester.

We run two sets of models: i) disc wind models; and ii) scale-free models.  In the disc wind models the rotation option in {\sc zeus2d} is turned on, introducing a rotational (centrifugal) pseudo-force. We include gravitational accelerations due to a point mass (of mass $M_*$) at the origin, and the base cells are given Keplerian velocities in the orbital direction. In the scale-free models both centrifugal and gravitational accelerations are turned off; these runs should therefore exactly match the  self-similar solutions . 

%%%%%%%%%%%%%%%%%%%%%%%%

\section{2D hydrodynamical simulations: comparison with  self-similar solutions}
\subsection{Comparison between  self-similar  solution and scale-free hydrodynamical simulations}

 The purpose of the scale-free models is to test our numerical method against the  self-similar solution. Figure 2 demonstrates that there is almost perfect agreement
between the  self-similar  solution (red) and scale-free hydrodynamic models
(black) for three values of $b$ between $0.75$ and $1.5$. For each hydrodynamic simulation we plot a series of re-scaled streamlines originating from different values of $R$. In each case the sonic point is found to lie within one grid cell of its position in the  self-similar  solution, and the excellent agreement between streamlines originating from different radii indicates that the numerical calculations are indeed scale-free. However, some small departures from self-similarity are visible in Figure 2 (particularly for $b=1.5$). These are due to the boundary conditions (which are by construction not scale-free) and other numerical effects, which we detail below.

The boundary conditions introduce two different numerical artefacts. First, the standard {\sc zeus} ``outflow'' boundary condition is exact only for supersonic flow along grid-lines (i.e., perpendicular to the boundary; Stone \& Norman 1992). As the flow is not purely radial, we invariably see some spurious reflection from the radial boundaries. This primarily occurs at the outer boundary, and is most prominent in the simulations with smaller values of $b$ (where the tangential velocity at the boundary is largest). This effect is most visible in Fig.\,4, where we see that the otherwise-constant launch velocity in the scale-free simulations increases progressively for $R/R_{\mathrm g} \gtrsim 8$. Test calculations with a larger outer grid radius ($R/R_{\mathrm g} = 20$) confirm that this is indeed a boundary effect, which alters the flow solution in the outer $\sim$20\% of the computational domain (see also discussion in Alexander et al.~2006).
  
A second artefact arises because the imposed base density profiles imply a radial pressure gradient for $b\ne0$, and are therefore not strictly consistent with the $v_r=0$ midplane boundary condition.  This effect is small in the scale-free simulations (and negligible in the disc wind simulations), but becomes more pronounced for larger values of $b$ and is the origin of the small departures from self-similarity seen in Fig.\,2 for $b=1.5$. Values of $b\gtrsim2$ result in simulations that show significant departures from self-similarity.
  
Finally, in the scale-free simulations (only) the required numerical resolution is not independent of $b$.  Smaller values of $b$ result in higher launch velocities, and the launch velocity approaches the sound speed for $b\lesssim0.5$. In such cases the sonic transition is poorly resolved, with the sonic point found only a few grid cells along each streamline. For $b\gtrsim0.6$ our calculations are well resolved, but for lower values of $b$ the sonic point is very close to the base of the flow, and the resolution required to achieve numerical convergence is prohibitively expensive. Given these numerical limitations, we restrict our hydrodynamic simulations to the range $b=0.75$--1.5\footnote{ Note in the disc wind runs the flow is accelerated over a length-scale $\sim R_{\mathrm g}$, which is always well resolved in our simulations. The lack of numerical convergence for small values of $b$ only occurs in the scale-free simulations.}

\subsection{Comparison between the scale-free/ self-similar solutions and the
disc wind simulations.}

\begin{figure}
\resizebox{\hsize}{!}{
\includegraphics[angle=270]{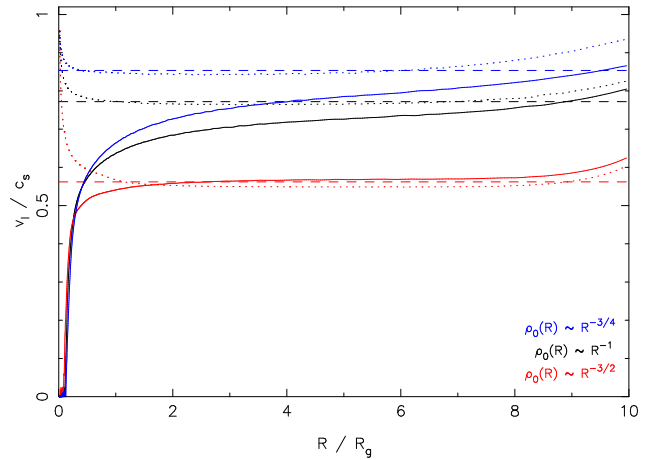}
}
\caption{The launch velocity at the flow base as a function of streamline
radius for the  self-similar solution (dashed), scale free hydrodynamical simulation
(dotted) and disc wind solutions (solid) for $b=0.75$ (blue), $b=1$ (black) and $b=1.5$
(red). The mild deviation of the scale free hydrodynamical solution
from constant launch velocity near the boundaries is a numerical artefact.
The rough constancy of the launch velocity for the disc wind solution
over a large dynamic range
%in the case $b=1$ and $b=1.5$ 
demonstrates the applicability of the
scale free approximation in these cases.}
% For $b=0.75$, the launch
%velocity rises with radius throughout the computational domain and
%the scale free condition is not attained.}
\end{figure}

 The disc wind models differ from those discussed above in that they include
rotation and gravitational acceleration by the central star. We expect such
models to approach the scale-free results in the limit of large $R/R_g$, but
here investigate the region over which the scale-free results are approximately
applicable to real disc winds.  We focus our comparisons on the streamline topology, as this uniquely determines both the launch velocity and (as the base density is fixed) the mass-loss profile.
Figure 3 compares the disc wind streamlines  and sonic surfaces 
with the scale free trajectories for $b$ in the range $0.75$ to $1.5$.  We
depict streamlines with base radii in the range $0.5 R_g$ to  $7.5 R_g$.
Note that whereas in the presence of gravity and rotation, the  wind velocity
drops steeply at small radii (so that there is an `inner most streamline'
at $\sim 0.2 R_g$; Font et al 2004), the scale free simulations naturally extend to arbitrarily
small radii. For the power law profiles considered here, the total mass
loss rate still converges at small radius but we caution that the
scale free solutions may over-estimate the signature generated by high
density wind tracers at small radii. 

  Quantitatively, the  self-similar solution places the sonic point very close to its true location for streamlines originating at $R \gtrsim R_g$.  In terms of the distance along each streamline, for $b=1.5$ we find that the sonic point in the scale-free solution is within $\pm3$\% of its location in the disc wind simulations for $R > R_g$, and is only 6.5\% in error even for the streamline originating at $0.5R_g$.  For $b=1.0$ the self-similar solution under-estimates the distance to the sonic point by $\simeq$5\% over most of the computational domain, but this rises to 13\% for the streamline originating at $R_g$, and 26\% at $0.5R_g$.  The solution for $b=0.75$ shows the least good fit with the disc wind model, but even then the agreement is very encouraging: the distance to the sonic point is within 15\% of the correct value over most of the domain, but is too small by 30\% at $R_g$ and 43\% at $0.5R_g$. As an additional test we also compute (integrated) mass-loss rates over the range $[0.5R_g,5R_g]$. As expected the scale-free solution over-estimates the mass-loss at small radii, but the agreement is still remarkably good: the ratios between the  self-similar  mass-loss rates and those in the disc wind simulation are 1.02, 1.13 and 1.16, for $b=1.5$, 1.0 \& 0.75 respectively.
 
 In general, the agreement between disc wind simulations and the scale free
solution improves at larger values of $R/R_g$ as expected. Nevertheless
the degree of agreement varies with $b$ in a way that can be simply understood
in terms of the curvature of the streamlines in the scale free solutions.
In the scale free case, the local streamline curvature is set by equation
(10) which balances the component of the pressure force normal to the
streamline with the convective derivative of the velocity. The solution
will thus be approximately scale free in cases where the component of
gravitational acceleration normal to the streamline is much less than the
convective derivative, 
  i.e. $u^2/R_{eff}  >> GM/R^2$. Given that the flow velocity
is of order $c_s$, this condition becomes $ (R/R_g)\times (R/R_{eff}) >>1 $.
Inspection of Figure 3 confirms this condition. In regions  where the scale
free solution has a small radius of curvature ($R_{eff} \le R$), the scale
free solutions provide a good match to the full disc wind solutions even at
relatively low values of $R/R_g$. This is particularly evident in the
case $b=1$ and also near the flow base in the case $b=1.5$. The agreement
in the case $b=1$, even at base radius as low as $R_b=0.5R_g$ is striking. The disc
wind solutions however deviate more strongly from the scale free solutions in
regimes where the streamlines are nearly straight (i.e. large $R_{eff}$). 
Such mild curvature is seen in the streamlines for the  $b=0.75$ case,
even at the flow base,  as a result
of the relatively weak pressure gradient in this case.  Mild curvature in the
scale free solution is also seen at larger heights in the $b=1.5$ case. This
contrasts with the $b=1$ case where the component of the pressure gradient 
normal to the streamlines changes sign over a short distance around $z/R_B \sim 1$:
at larger heights the streamlines are concave upwards because the pressure declines
with increasing height. For $b=1.5$, by contrast, the pressure gradient at large
heights is
small in magnitude and the streamlines are almost straight. In all cases where
the scale free solution yields solutions with mild curvature, the addition
of gravity modifies the streamlines, yielding solutions that are  concave  upwards.

 In summary, the scale free solutions do a remarkably good job at approximating
the disc wind solutions for $b=1$ and $b=1.5$ although there is some
deviation in the latter case for base radii within a few times $R_g$. Even this
latter deviation is however only apparent at heights $ z > R_b$; the good agreement
near the flow base means that the launch velocities are independent
of streamline (as in the scale free solution) even for $R/R_g$ as low as $1$
(see Figure 4). In the case of $b=0.75$, by contrast, the scale free
solution exhibits mild curvature throughout and thus gravity plays an important
role in setting the streamline topology even at $R/R_g$ as large as $10$. This
is also demonstrated by Figure 4, which shows that for $b=0.75$ the disc wind solutions
never attain the limit of constant launch velocity (as required by a
scale-free solution) within the computational grid ($R/R_g < 10$).

\section{Conclusions}

 We have developed a similarity solution for the structure of an isothermal
disc wind with a power law base density profile ($\rho \propto R_b^{-b}$).
The problem is strictly scale
free only in the case that both rotation and gravity are neglected; we have verified
that the solutions obtained are in excellent agreement with hydrodynamic simulations
in this case and that the streamline shape becomes progressively more vertical
as $b$ (the index of the base density power law) is reduced. The results can be simply
understood in terms of the force balance perpendicular to the streamlines which
implies that streamlines become more curved for steeper density profiles (see Figure 2).

 We have also compared these solutions with disc wind simulations which also
include Keplerian rotation and the gravity of the central object. We find that
the  self-similar  solution provides a good match to the disc wind simulations over
a wide range of radii.  This agreement is particularly good in 
the case of  
%, apart from in the  case of the shallowest density profile ($b=0.75$)
%where gravity plays an important role in shaping the streamline topology even at large
%radii and where a scale free solution is not obtained. 
%In the case of 
the steeper
profiles ($b=1$ and $b=1.5$, which are more appropriate to those expected in
photoevaporating winds; Font et al 2004).
%, the {\bf self-similar}  solution provides a remarkably
%good approximation to the disc wind solution. 
In the case $b=1$ this excellent agreement
extends in to streamlines originating from a factor two within $R_g$ (see Figure 3).  

   The  self-similar  solution derived here will be  useful for the  
 modelling of  disc winds without  recourse to hydrodynamic simulations. 
There are numerous potential applications in terms of modeling the line profiles
and free-free emission from thermally driven disc winds, particularly
in the protoplanetary disc context. Such solutions also provide a useful tool
for benchmarking simulations involving the entrainment of dust by disc winds
(cf Hutchison \& Laibe 2016).

\section{Acknowledgments}
We thank James Owen for useful discussions  and the referee for comments that have helped to
improve the paper's clarity. This work has been partially
supported  by  the DISCSIM project, grant agreement 341137 funded by the European Research Council under ERC-2013-ADG. RDA acknowledges support from STFC through an Advanced Fellowship (ST/G00711X/1), and from the Leverhulme Trust through a Philip Leverhulme Prize. Astrophysical research at the University of Leicester is supported by an STFC Consolidated Grant (ST/K001000/1).  This research used the ALICE High Performance Computing Facility at the University of Leicester. Some resources on ALICE form part of the DiRAC Facility jointly funded by STFC and the Large Facilities Capital Fund of BIS.  This work also used the DiRAC {\it Complexity} system, operated by the University of Leicester IT Services, which forms part of the STFC DiRAC HPC Facility ({\tt http://www.dirac.ac.uk}). This equipment is funded by BIS National E-Infrastructure capital grant ST/K000373/1 and  STFC DiRAC Operations grant ST/K0003259/1. DiRAC is part of the UK National E-Infrastructure.

{}

\section{Appendix: Derivation of the convective derivative}

 The streamline geometry is set by a requirement of hydrodynamic force
balance perpendicular to the flow streamlines wherein the component
of the acceleration due to the pressure gradient in this direction is
matched by the corresponding component of the convective derivative,
$u.\nabla u$. As in the main text, we denote unit vectors perpendicular
and parallel to the streamline by $\hat l$ and $\hat s$ respectively. 
Here we will show that $(u.\nabla u) . \hat l = -u^2/R_{eff}$ (see equation
10) where $R_{eff}$ is the local radius of curvature of the streamline
such that $R_{eff} > 0$ implies that the streamline is convex upwards
(i.e. in the direection of increasing $l$).

 We consider a 2D coordinate system {s,l} where $l$ is the
perpendicular distance of any point P from a fixed (reference) streamline
which passes through point O (coordinates {0,0}) and where $s$ is the
distance measured along the reference streamline between point O and
the point on the streamline whose normal passes through P. Consider now
points A and B with coordinates {0,l} and
{ds, l+dl}. If the radius of curvature of the streamline at
0 is $R_{eff}$  then the distance between points A and B can be written: 

\begin{equation}
AB^2 = dl^2 + \biggl({{R_{eff}+l}\over{R_{eff}}}\biggr)^2 ds^2
\end{equation}

The components of the metric tensor in this coordinate
system are thus $g_{ll} =1$ and $g_{ss}= \biggl({{R_{eff}+l}\over{R_{eff}}}\biggr)^2$.

 The definition of the convective derivative with respect to
arbitrary coordinates $q_i$ is given (e.g. http://mathworld.wolfram.com/ConvectiveOperator.html) by 

\begin{equation}
 [u.\nabla u]_j = \Sigma_{k=1}^{k=2} \biggl({{u_k}\over{h_k}} {{\partial u_j}\over{\partial q_k}} + {{u_k}\over{h_k h_j}} \bigl(u_j {{\partial h_j}\over{\partial q_k}} - u_k {{\partial h_k}\over{\partial q_j}}\bigr) \biggr)
\end{equation}

\noindent where $h_i^2 =  g_{ii}$.  

 Since the coordinate $s$ lies along the streamline direction,
we have $u_s=  u$
and $u_l=0$; this implies:

\begin{equation}
 [u.\nabla u]_l = -{{u}\over{h_s h_l}}u {{\partial h_s}\over{\partial l}}
\end{equation} 

\noindent i.e.
\begin{equation}
 [u.\nabla u]_l = -u^2 {{\partial {\rm{ln}} h_s}\over{\partial l}} 
\end{equation}

\noindent Since {\footnote{Note that in deriving the identity equation (28) we are
considering the component of $u.\nabla u$ at an arbitary point O and define a
coordinate system based on the  streamline passing through O with a particular value of $R_{eff}$. For
this derivation, $R_{eff}$ is then a {\it fixed} property of the coordinate system and
is not a function of $l$. The derived identity is then valid at all points, regardless
of whether, in a given velocity field, $R_{eff}$ varies between streamlines.}}
 
\begin{equation}
{{\partial {\rm{ln}} h_s}\over{\partial l}} = {{1}\over{R_{eff} + l}}
\end{equation}

\noindent then at point O ($l=0$), this is simply $1/R_{eff}$.
Thus
\begin{equation}
[u.\nabla u]_l = -{{u^2}\over{R_{eff}}}
\end{equation}   
\label{lastpage}
\end{document}